%% file: main.tex
\documentclass[conference]{IEEEtran}

\usepackage{cite}
\usepackage{graphicx}
\usepackage{booktabs}
\usepackage{tabularx}
\usepackage{array}
\usepackage{url}
\usepackage{multirow}
\usepackage{tikz}
\usetikzlibrary{positioning,arrows.meta,fit,backgrounds,calc}

\newcolumntype{Y}{>{\raggedright\arraybackslash}X}
\newcommand{\system}{\textsc{IndustriConnect}}

\InputIfFileExists{generated/tables/exact_counts.tex}{}{C
  \newcommand{\totalNormalRuns}{480}
  \newcommand{\totalFaultRuns}{210}
  \newcommand{\totalStressRuns}{120}
  \newcommand{\totalRecoveryTrials}{60}
  \newcommand{\totalRuns}{870}
  \newcommand{\totalToolCalls}{0}
  \newcommand{\normalToolCalls}{0}
  \newcommand{\faultToolCalls}{0}
  
  \newcommand{\modbusToolCount}{20}
  \newcommand{\mqttToolCount}{15}
  \newcommand{\opcuaToolCount}{7}
}

\begin{document}

\title{IndustriConnect: MCP Adapters and Mock-First Evaluation for AI-Assisted Industrial Operations}

\author{
\IEEEauthorblockN{Melwin Xavier}
\IEEEauthorblockA{Lule\aa{} tekniska universitet, Sweden\\
melwin.xavier@ltu.se}
\and
\IEEEauthorblockN{Melveena Jolly}
\IEEEauthorblockA{Independent Researcher\\
melveenajollyk@gmail.com}
\and
\IEEEauthorblockN{Vaisakh M A}
\IEEEauthorblockA{Independent Researcher\\
997vaisakh@gmail.com}
\and
\IEEEauthorblockN{Midhun Xavier}
\IEEEauthorblockA{Independent Researcher\\
midhun@industriagents.com}
}

\maketitle

\begin{abstract}
AI assistants can decompose multi-step workflows, but they do not natively speak industrial protocols such as Modbus, MQTT/Sparkplug B, or OPC UA. This paper presents \system, a prototype suite of Model Context Protocol (MCP) adapters that expose industrial operations as schema-discoverable AI tools while preserving protocol-specific connectivity and safety controls. The system uses a common response envelope and a mock-first workflow so adapter behavior can be exercised locally before connecting to plant equipment. A deterministic benchmark covering normal, fault-injected, stress, and recovery scenarios evaluates the flagship adapters. The benchmark comprises \totalRuns{} runs (\totalNormalRuns{} normal, \totalFaultRuns{} fault-injected, \totalStressRuns{} stress, \totalRecoveryTrials{} recovery trials) and \totalToolCalls{} tool calls across 7 fault scenarios and 12 stress scenarios. The normal suite achieved full success, the fault suite confirmed structured error handling with adapter-level uint16 range validation, and the stress suite identified concurrency boundaries. Same-session recovery after endpoint restart is demonstrated for all three protocols. The results provide evidence spanning adapter correctness, concurrency behavior, and structured error handling for AI-assisted industrial operations.
\end{abstract}

\section{Introduction}
\label{sec:introduction}

Consider a simple operator request: check a temperature excursion on an OPC UA namespace, confirm a related Modbus register block, and publish a Sparkplug B status update for downstream systems. Each subtask is routine in isolation, but the overall workflow crosses incompatible operational technology (OT) protocols whose interfaces were not designed for modern AI assistants. This protocol gap remains a practical barrier to AI-assisted industrial automation even as Industry 4.0 programs push tighter coupling between information technology and plant-floor systems \cite{xu2018industry4,wollschlaeger2017future}.

Large language model (LLM) assistants work best with typed, discoverable tool interfaces rather than binary protocol frames or vendor SDKs. MCP provides a standard way for assistants to list tools, inspect schemas, and invoke operations through a common call interface \cite{anthropic2024mcp}. MCP alone, however, does not solve OT connectivity, endpoint health, or the need to represent blocked writes and transient failures in a way that an agent can reason about safely.

That gap is especially visible in brownfield environments. Real plants often begin with partial register maps, vendor-specific addressing rules, or limited access windows on production equipment. A practical integration layer therefore needs more than tool schemas: it also needs protocol adapters, explicit safety conventions, and a way to validate workflows before touching live controllers.

This paper reframes \system\ as a bounded systems contribution: a prototype MCP-to-OT adapter architecture with a mock-first evaluation workflow. The paper does not claim a complete industrial platform, a new LLM reasoning method, or production deployment evidence. Instead, it focuses on the flagship adapters for Modbus, MQTT/Sparkplug B, and OPC UA while retaining seven additional protocol modules as ecosystem breadth.

The manuscript makes three contributions:
\begin{enumerate}
    \item A practical MCP-to-OT adapter architecture that standardizes tool discovery and response handling across heterogeneous industrial protocols.
    \item A mock-first methodology for validating adapter behavior, write guards, and restart recovery, extended with fault injection and stress testing to expose error-handling and concurrency boundaries without requiring physical devices.
    \item A reproducible deterministic evaluation on Modbus, MQTT/Sparkplug B, and OPC UA comprising normal, fault-injected, stress, and recovery suites using deterministic workloads with statistical variance reporting.
\end{enumerate}

The evaluation uses a deterministic benchmark that exercises adapter behavior under normal operation, deliberate fault conditions, stress scenarios, and endpoint restart.

\section{Design Goals and Architecture}
\label{sec:architecture}

\textbf{Heterogeneity.} Industrial protocols expose different interaction models: register maps for Modbus, broker topics for MQTT, and browsable information models for OPC UA \cite{cavalieri2013opcua,mishra2020mqtt}. The adapter layer must translate these differences into tool schemas that remain understandable to an AI client.

\textbf{Schema discoverability.} Each adapter registers MCP tools so a client can enumerate available operations and call them without bespoke protocol code. This keeps the assistant-facing interface uniform even when the underlying transport or addressing model differs.

\textbf{Write safety.} The adapters expose explicit write operations, but writes remain bounded by per-server configuration such as disable flags, address validation, and structured error returns. These controls matter because an AI-facing API must represent both successful writes and intended denials in a machine-readable way.

\textbf{Mockability.} Every flagship adapter is paired with a local mock so the full tool path can be exercised offline. This enables reproducible evaluation, debugging, and regression checks before any real controller or fieldbus is involved.

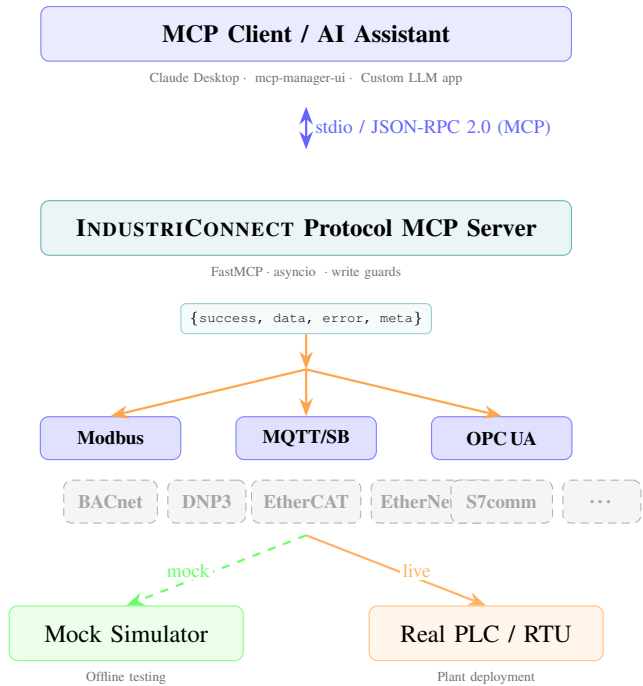
\begin{figure}[t]
    \centering
    \resizebox{0.95\columnwidth}{!}{%
    \begin{tikzpicture}[
        >=Stealth,
        box/.style={draw, rounded corners=3pt, minimum height=0.7cm,
                     text centered, font=\small},
        layer/.style={box, minimum width=6.8cm, font=\small\bfseries},
        proto/.style={box, minimum width=1.8cm, minimum height=0.55cm,
                      font=\scriptsize\bfseries},
        scaffold/.style={proto, draw=gray!60, fill=gray!8, text=gray!70,
                         densely dashed},
        envfield/.style={font=\tiny\ttfamily, text=black!70},
        arrlab/.style={font=\scriptsize, fill=white, inner sep=1pt},
    ]
    % --- Layer 1: AI Assistant ---
    \node[layer, fill=blue!8, draw=blue!50] (ai)
        {MCP Client / AI Assistant};
    \node[font=\tiny, text=black!55, below=0.02cm of ai]
        (aisub) {Claude Desktop\;\textperiodcentered\;
                  mcp-manager-ui\;\textperiodcentered\;
                  Custom LLM app};

    % --- Arrow 1 ---
    \draw[<->, thick, blue!60]
        ($(aisub.south)+(0,-0.15)$) -- ++(0,-0.55)
        node[arrlab, midway, right=2pt] {stdio / JSON-RPC 2.0 (MCP)};

    % --- Layer 2: Adapter ---
    \node[layer, fill=teal!8, draw=teal!60,
          below=1.35cm of aisub] (adapter)
        {\system{} Protocol MCP Server};
    \node[font=\tiny, text=black!55, below=0.02cm of adapter]
        (adsub) {FastMCP\;\textperiodcentered\;asyncio
                 \;\textperiodcentered\;write guards};

    % --- Envelope ---
    \node[draw=teal!40, rounded corners=2pt, fill=teal!4,
          inner sep=3pt, font=\tiny,
          below=0.18cm of adsub] (env)
        {\texttt{%
          \{success, data, error, meta\}}};

    % --- Arrow 2 (splits to 3 protocols) ---
    \coordinate (split) at ($(env.south)+(0,-0.45)$);
    \draw[->, thick, orange!70] (env.south) -- (split);

    % --- Protocol nodes ---
    \node[proto, fill=blue!12, draw=blue!50,
          below left=0.6cm and 1.6cm of split] (modbus) {Modbus};
    \node[proto, fill=blue!12, draw=blue!50,
          below=0.6cm of split] (mqtt) {MQTT/SB};
    \node[proto, fill=blue!12, draw=blue!50,
          below right=0.6cm and 1.6cm of split] (opcua) {OPC\,UA};

    \draw[->, thick, orange!70] (split) -- (modbus.north);
    \draw[->, thick, orange!70] (split) -- (mqtt.north);
    \draw[->, thick, orange!70] (split) -- (opcua.north);

    % --- Scaffold badges ---
    \node[scaffold, below=0.25cm of modbus, minimum width=1.2cm]
        (s1) {BACnet};
    \node[scaffold, right=0.12cm of s1, minimum width=1.0cm]
        (s2) {DNP3};
    \node[scaffold, below=0.25cm of mqtt, minimum width=1.4cm]
        (s3) {EtherCAT};
    \node[scaffold, right=0.12cm of s3, minimum width=1.5cm]
        (s4) {EtherNet/IP};
    \node[scaffold, below=0.25cm of opcua, minimum width=1.3cm]
        (s5) {S7comm};
    \node[scaffold, right=0.12cm of s5, minimum width=1.0cm]
        (s6) {\dots};

    % --- Layer 3: Device targets ---
    \coordinate (devtop) at ($(s3.south)+(0,-0.4)$);
    \node[box, fill=green!8, draw=green!50, minimum width=3.0cm,
          below left=0.65cm and 0.8cm of devtop] (mock)
        {Mock Simulator};
    \node[box, fill=orange!8, draw=orange!50, minimum width=3.0cm,
          below right=0.65cm and 0.8cm of devtop] (real)
        {Real PLC / RTU};
    \node[font=\tiny, text=black!55, below=0.01cm of mock]
        {Offline testing};
    \node[font=\tiny, text=black!55, below=0.01cm of real]
        {Plant deployment};

    \draw[->, thick, green!60, dashed]
        ($(s3.south)+(0,-0.15)$) -- (mock.north)
        node[arrlab, midway, left=1pt] {\scriptsize mock};
    \draw[->, thick, orange!60]
        ($(s3.south)+(0,-0.15)$) -- (real.north)
        node[arrlab, midway, right=1pt] {\scriptsize live};

    \end{tikzpicture}%
    }
    \caption{\system\ adapter architecture. An AI assistant issues MCP tool calls through a common JSON-RPC interface. The adapter layer handles protocol translation and safety guards, returning a shared response envelope. Each protocol adapter targets either a local mock (dashed) or real plant equipment.}
    \label{fig:architecture}
\end{figure}

Figure \ref{fig:architecture} shows the core pattern. An MCP-compatible assistant issues tool calls to an \system\ adapter. The adapter handles protocol translation, connection management, and safety guards, then talks to either a mock endpoint or plant equipment. All benchmarked adapters return the same response envelope,
\texttt{\{success, data, error, meta\}}. The \texttt{success} flag indicates whether the requested action achieved the expected protocol outcome, \texttt{data} holds the protocol-specific payload, \texttt{error} carries a structured failure reason, and \texttt{meta} records endpoint details, timings, and retry counts. This shared envelope does not erase protocol semantics; it defines a common measurement boundary and lets cross-protocol orchestration observe healthy reads, blocked writes, and reconnect behavior with the same outer structure.

\begin{table}[t]
    \caption{Shared response-envelope fields used across the benchmarked adapters.}
    \label{tab:envelope}
    \centering
    \scriptsize
    \begin{tabularx}{\columnwidth}{@{}lY@{}}
        \toprule
        Field & Role in the adapter contract \\
        \midrule
        \texttt{success} & Reports whether the requested protocol action reached the expected outcome. \\
        \texttt{data} & Carries the protocol-specific payload such as registers, node values, or publish metadata. \\
        \texttt{error} & Preserves a machine-readable failure reason, including guarded denials and transient endpoint failures. \\
        \texttt{meta} & Records wall-clock latency, endpoint details, attempts, and protocol-specific trace context. \\
        \bottomrule
    \end{tabularx}
\end{table}

Table \ref{tab:envelope} is operationally important because it defines how the benchmark judges behavior. A blocked write is not treated as a generic failure if the task oracle expects a guarded denial, while a reconnect path is evaluated through the same outer structure as a healthy read. This is the main reason the paper can compare three different protocol families without pretending they expose the same payload model.

The mock-first workflow complements the same architecture. Each flagship adapter is exercised against a local endpoint first, but the MCP layer remains identical when the endpoint changes. That lets an engineer develop prompts, schemas, and safety checks against a deterministic mock while preserving the same tool surface for later hardware validation.

\section{Prototype Implementation}
\label{sec:implementation}

Every protocol module follows the same skeleton: an MCP server registers tools and schemas, a protocol client wrapper translates tool calls into native operations, and a mock endpoint provides a safe development target. The benchmarked Modbus, MQTT/Sparkplug B, and OPC UA stacks all use this pattern, including the shared response envelope and environment-driven write gating. The result is not a universal industrial abstraction layer; it is a consistent adapter contract around heterogeneous protocols.

At implementation level, the common skeleton yields four reusable behaviors. First, each server exposes discoverable MCP tools with bounded JSON input shapes. Second, each adapter keeps transport- and protocol-specific state in its own client wrapper rather than leaking it into the LLM-facing layer. Third, every tool normalizes timing and error details into the shared response envelope. Fourth, each protocol stack ships with a mock so smoke tests, prompt experiments, and regression checks can run without physical hardware.

\begin{table}[t]
    \caption{Repository protocol inventory. Tool counts come from registered handlers in the benchmark pass.\protect\footnotemark}
    \label{tab:inventory}
    \centering
    \scriptsize
    \setlength{\tabcolsep}{2pt}
    \begin{tabularx}{\columnwidth}{@{}p{0.78in}cp{0.72in}p{0.6in}p{0.82in}@{}}
        \toprule
        Protocol & Tools & Mock & Status & Rep. operation \\
        \midrule
        Modbus & \modbusToolCount{} & TCP device & flagship & register/coil I/O \\
        MQTT + SB & \mqttToolCount{} & broker sim. & flagship & pub/sub + DDATA \\
        OPC UA & \opcuaToolCount{} & UA server & flagship & browse + node R/W \\
        \midrule
        BACnet/IP & 7 & BA device & scaffold & object properties \\
        DNP3 & 8 & outstation & scaffold & point poll/control \\
        EtherCAT & 14 & slave & scaffold & PDO/SDO access \\
        EtherNet/IP & 17 & PLC & scaffold & controller tags \\
        PROFIBUS DP/PA & 11 & slave & scaffold & bus scan + cyclic I/O \\
        PROFINET & 14 & IO device & scaffold & discovery + IO exchange \\
        Siemens S7comm & 20 & PLC & scaffold & DB/I/O diagnostics \\
        \bottomrule
    \end{tabularx}
\end{table}
\footnotetext{Protocols below the mid-rule are not evaluated in this paper.}

Table \ref{tab:inventory} keeps the breadth story honest. The repository contains ten protocol modules, but only three are benchmarked in this paper. The Modbus adapter exposes \modbusToolCount{} tools; the benchmark evaluates a representative subset covering connectivity, reads, writes, guarded denials, and fault paths. The remaining seven modules remain useful context because they show that the MCP adapter pattern generalizes across fieldbus, Ethernet, and broker-centric industrial stacks even when several implementations are still roadmap-driven.

That breadth-versus-evidence split is deliberate. The three flagship stacks were chosen because they represent different industrial interaction styles and because their current mocks are stable enough for repeated quantitative runs. Modbus covers direct register-style access, MQTT/Sparkplug B covers broker-mediated messaging, and OPC UA covers hierarchical information-model access. The remaining stacks are better understood as evidence that the repository architecture generalizes, not as equally validated research claims.

\textbf{Flagship adapters.} The Modbus adapter centers on register and coil operations, connection health checks, guarded writes, and uint16 range validation. The MQTT adapter exposes broker inspection, subscriptions, generic publishes, and Sparkplug B device-data publishes. The OPC UA adapter focuses on node browsing, node reads, writes, multi-node access, and reconnection logic that re-establishes connections after endpoint restarts. Method invocation exists in the current OPC UA prototype but was excluded from the benchmark because the current mock method path is not yet stable enough for repeatable evaluation.

\textbf{Modbus details.} The current Modbus surface contains \modbusToolCount{} tools, including register and coil reads, bulk writes, typed holding-register access, masked register writes, device-information queries, and alias-based access paths. The adapter validates uint16 range boundaries (0--65535) before forwarding write operations to the protocol layer, ensuring overflow values are rejected with a structured error rather than silently truncated. The benchmark intentionally concentrates on the core cases most likely to appear in assistant workflows: connectivity checks, block reads, readback verification after a write, and a write-disabled safety path.

\textbf{MQTT/Sparkplug B details.} The MQTT adapter exposes \mqttToolCount{} tools spanning broker status, topic subscriptions, generic publish/unsubscribe operations, and Sparkplug-specific birth, death, data, and command flows. In practice this makes the adapter suitable both for ordinary pub/sub workflows and for structured industrial telemetry updates. The benchmark therefore mixes a broker inspection task, a wildcard subscription task, a plain control publish, and a Sparkplug B DDATA publish rather than evaluating only one messaging style.

\textbf{OPC UA details.} The OPC UA surface is smaller at \opcuaToolCount{} tools, but the semantics are richer because browsing and variable enumeration traverse an information model rather than a flat address map. The adapter includes reconnection logic: a liveness probe checks the ServerStatus node before each operation, and on failure the stale client is disconnected and a fresh connection is established. The benchmark covers point reads, browsing, writes with readback confirmation, and full variable discovery over the mock plant.

\textbf{Mock endpoint fidelity.} Each flagship mock simulates enough plant-floor state for the benchmark workload. The Modbus mock provisions 100 holding registers (valve\_position, heater\_power, fan\_speed, conveyor\_speed, command words), 100 input registers (temperature, pressure, flow\_rate, tank\_level, vibration, pH, humidity, motor\_speed, production counters), 100 coils, and 100 discrete inputs, with a 1\,Hz simulation loop that models inter-variable influence (heater power affects temperature, pump state affects pressure and flow). The MQTT mock combines an aedes broker with a Sparkplug B edge-node simulator that publishes NBIRTH, DBIRTH, and periodic DDATA messages for two devices; device metrics follow sinusoidal variation and all Sparkplug payloads use protobuf encoding. The OPC UA mock exposes 8 read-only sensor variables, 6 read-write actuator variables, system-status variables, and 5 callable methods under a single namespace, updated every second with state-driven dynamics mirroring the Modbus model.

\begin{figure}[t]
    \centering
    \includegraphics[width=0.92\columnwidth]{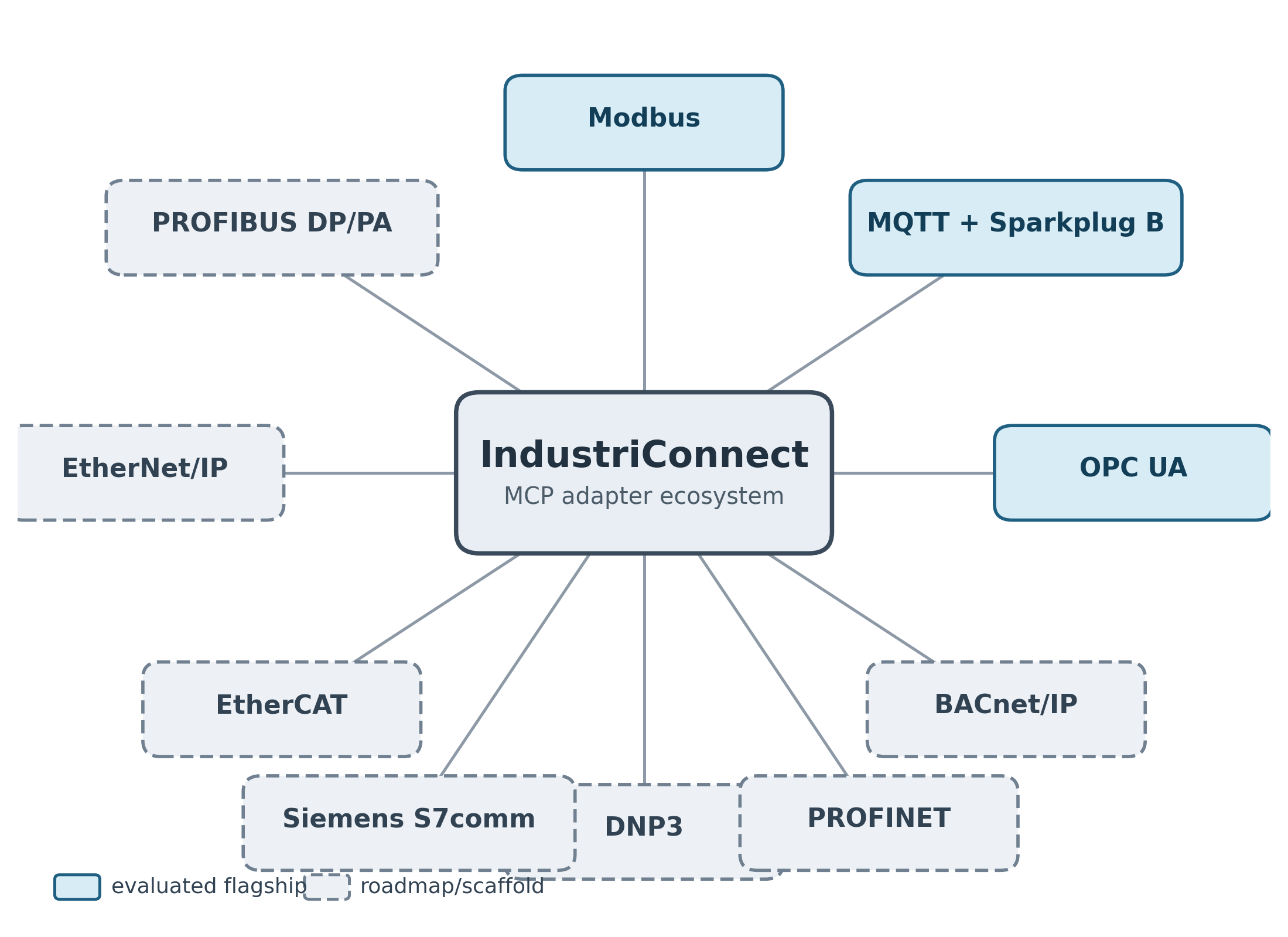}
    \caption{Repository protocol landscape with the three evaluated flagship adapters highlighted. This figure is supporting context, not evaluation evidence.}
    \label{fig:protocolmap}
\end{figure}

\begin{figure}[t]
    \centering
    \includegraphics[width=0.92\columnwidth]{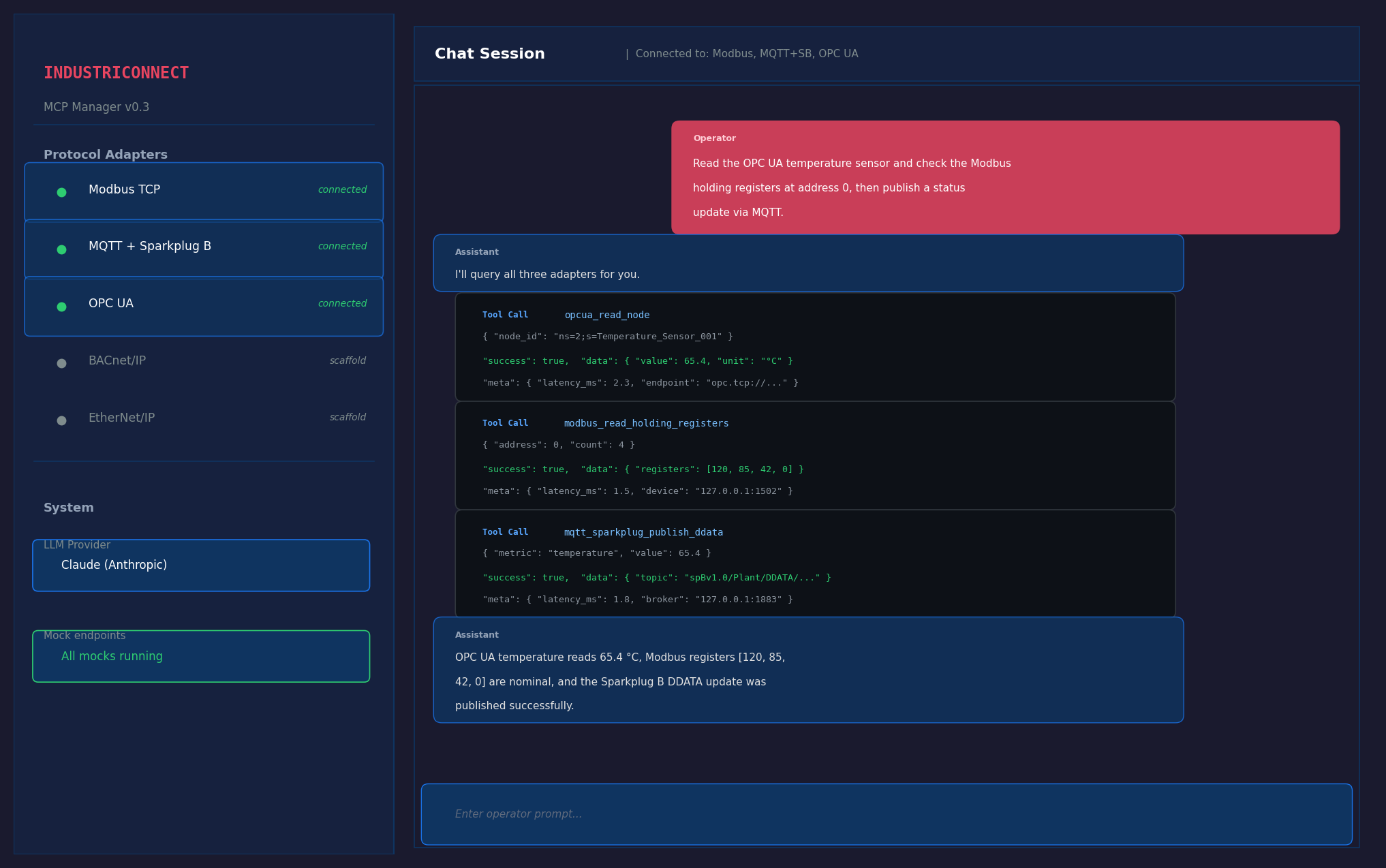}
    \caption{MCP Manager UI used to register adapters and issue operator-style prompts. This figure is supporting context, not benchmark evidence.}
    \label{fig:managerui}
\end{figure}

\textbf{Supporting console.} The repository also includes a browser-based MCP Manager UI for registering protocol adapters and issuing operator-style prompts. Figure \ref{fig:protocolmap} shows the corrected ten-protocol landscape, and Figure \ref{fig:managerui} shows the MCP Manager UI. Both figures are supporting context only; they are not part of the benchmark and are not counted as separate research contributions.

\section{Experimental Setup}
\label{sec:setup}

The evaluation targets adapter behavior rather than end-to-end LLM planning. To keep the workload reproducible, the authors fixed representative assistant intents and executed them as deterministic MCP tool workflows against local mocks. This isolates connection handling, tool schemas, response normalization, and safety behavior from prompt variability \cite{yao2022react}. All measurements were taken on localhost against mock endpoints; latencies therefore reflect adapter overhead, not network round-trip times.

\begin{table}[t]
    \caption{Benchmark matrix for the deterministic evaluation.}
    \label{tab:tasks}
    \centering
    \scriptsize
    \begin{tabularx}{\columnwidth}{@{}llY@{}}
        \toprule
        Suite & Family & Task set \\
        \midrule
        \multirow{4}{*}{Normal} & Modbus & Ping; four-register read; register write/readback; guarded write denial \\
        & MQTT/SB & Broker info; subscribe to \texttt{sensors/\#}; generic publish; Sparkplug B DDATA publish \\
        & OPC UA & Temperature read; actuator browse; valve write/readback; full variable enumeration \\
        & Cross & Sequential snapshot (X1); sequential control (X2); parallel snapshot (X1p); parallel control (X2p) \\
        \midrule
        \multirow{4}{*}{Fault} & Modbus & Read invalid register (FM1); write overflow value---adapter rejects (FM2) \\
        & MQTT/SB & Publish to empty topic (FQ1); subscribe with invalid QoS (FQ2) \\
        & OPC UA & Read non-existent node (FO1); write wrong type to float node (FO2) \\
        & Cross & 3-step sequence with one deliberate bad input (FX1) \\
        \midrule
        \multirow{3}{*}{Stress} & Per-adapter & 4 concurrent reads (S1--S3); concurrent read+write (S4--S6) \\
        & Per-adapter & 50 sequential rapid-fire reads (S7--S9) \\
        & Per-adapter & Mid-operation mock restart (S10--S12) \\
        \bottomrule
    \end{tabularx}
\end{table}

Table \ref{tab:tasks} summarizes the workload. The normal suite contains 16 tasks (four Modbus, four MQTT/Sparkplug B, four OPC UA, four cross-protocol including parallel variants), each repeated 30 times for \totalNormalRuns{} runs. The fault-injection suite adds 7 tasks repeated 30 times each for \totalFaultRuns{} runs. The stress suite adds 12 tasks repeated 10 times each for \totalStressRuns{} runs. The recovery suite injects one transient outage per flagship stack, repeated 20 times each for \totalRecoveryTrials{} restart trials. The full benchmark thus comprises \totalRuns{} runs and \totalToolCalls{} tool calls. All aggregation includes mean, standard deviation, and 95\% confidence intervals (t-distribution).

The fault-injection philosophy follows established practice~\cite{natella2016faultinjection}: correctness under fault conditions means returning \texttt{success=false} with a meaningful, structured error envelope---not silently succeeding or crashing. A fault task passes when the adapter produces a well-formed error response that an assistant could interpret and act on. FM2 (Modbus overflow) now tests a deterministic boundary: the adapter validates uint16 range and must reject the value with \texttt{success=false}---the check is falsifiable.

The recovery suite uses same-session semantics: after restarting the mock endpoint, the harness first attempts recovery through the original MCP session, then falls back to a fresh session. This dual-mode approach distinguishes adapters that support transparent reconnection (e.g., pymodbus auto-reconnect, OPC UA liveness probe) from those requiring a new session.

\begin{table}[t]
    \caption{Benchmark runtime configuration used in the paper.}
    \label{tab:runtime}
    \centering
    \scriptsize
    \begin{tabularx}{\columnwidth}{@{}lYY@{}}
        \toprule
        Family & Mock endpoint & Recovery semantics \\
        \midrule
        Modbus & TCP mock at \texttt{127.0.0.1:1502} & Same-session via pymodbus auto-reconnect, then fresh session \\
        MQTT/SB & Local broker at \texttt{127.0.0.1:1883} & Same-session via paho \texttt{reconnect\_delay\_set}, then fresh session \\
        OPC UA & Local OPC UA endpoint & Same-session via \texttt{\_ensure\_connected()} liveness probe, then fresh session \\
        \bottomrule
    \end{tabularx}
\end{table}

All benchmarked adapters were launched as local stdio MCP servers. The Modbus stack targeted the TCP mock at \texttt{127.0.0.1:1502}, the MQTT adapter targeted the local broker at \texttt{mqtt://127.0.0.1:1883}, and the OPC UA adapter targeted the local mock endpoint at \texttt{opc.tcp://127.0.0.1:4840}. The benchmark harness opened dedicated sessions for the read-write Modbus configuration, the read-only Modbus configuration, MQTT, and OPC UA, then executed the normal, fault, stress, and recovery suites in a fixed order.

The normal runs also use fixed value patterns so task validation is deterministic. For example, the Modbus write/readback task writes \texttt{40 + repetition}, the Sparkplug task publishes a single temperature metric with \texttt{20.0 + repetition}, and the OPC UA valve task writes \texttt{25.0 + repetition}. The cross-protocol control task uses a fixed Modbus write, an OPC UA boolean write, and a control-topic publish, making the end-to-end sequence stable enough for repeated latency summaries.

The harness records raw tool responses and then derives family-level, task-level, and recovery-phase summaries from the same JSON artifacts. The tables in the paper are generated from these JSON results through a single-source-of-truth pipeline (\texttt{generate\_paper\_tables.py}), eliminating hand-copied numbers.

\smallskip
\noindent\textbf{Workflow example (X1).} An operator asks for a fast cross-system snapshot before escalating a maintenance event. The assistant issues three fixed tool calls: a Modbus sensor-block read, an OPC UA multi-node read, and an MQTT broker inspection. The resulting state bundle confirms that the transport links are healthy and that the key process values can be read through all three adapters in one short sequence. X1p exercises the same workflow with parallel tool calls via \texttt{asyncio.gather()}.
\smallskip

\section{Results}
\label{sec:results}

\subsection{Normal Suite Results}
\label{sec:results:normal}

\begin{table}[t]
    \caption{Family-level results from the normal benchmark suite (30 repetitions). Cross-protocol rows aggregate the four multi-adapter tasks.}
    \label{tab:results}
    \centering
    \scriptsize
    \setlength{\tabcolsep}{3pt}
    \input{generated/tables/tab_results}
\end{table}

Table \ref{tab:results} shows clean success rates across the normal suite: all \totalNormalRuns{} task executions succeeded, and all \normalToolCalls{} tool calls matched their expected outcomes. Modbus and MQTT/Sparkplug B remained in the low single-digit millisecond range. OPC UA shows wider spread due to variable enumeration, which is two orders of magnitude slower than single-node reads. Cross-protocol tasks remain close to 7\,ms median, suggesting the response-envelope approach does not impose noticeable overhead for short local workflows. All latencies reported are localhost measurements; network round-trip would dominate in a deployed setting.

\begin{table*}[t]
    \caption{Per-task latency breakdown from the normal benchmark suite (30 repetitions).}
    \label{tab:taskbreakdown}
    \centering
    \scriptsize
    \setlength{\tabcolsep}{4pt}
    \input{generated/tables/tab_taskbreakdown}
\end{table*}

Table \ref{tab:taskbreakdown} makes the tail behavior clearer. Modbus remains tightly clustered even when readback verification or guarded-write rejection is required. OPC UA variable enumeration dominates the family p95, consistent with information-model breadth rather than a generalized slowdown. The parallel cross-protocol variants (X1p, X2p) show reduced latency compared to their sequential counterparts (X1, X2), confirming that concurrent adapter invocation works correctly.

\begin{figure}[t]
    \centering
    \includegraphics[width=\columnwidth]{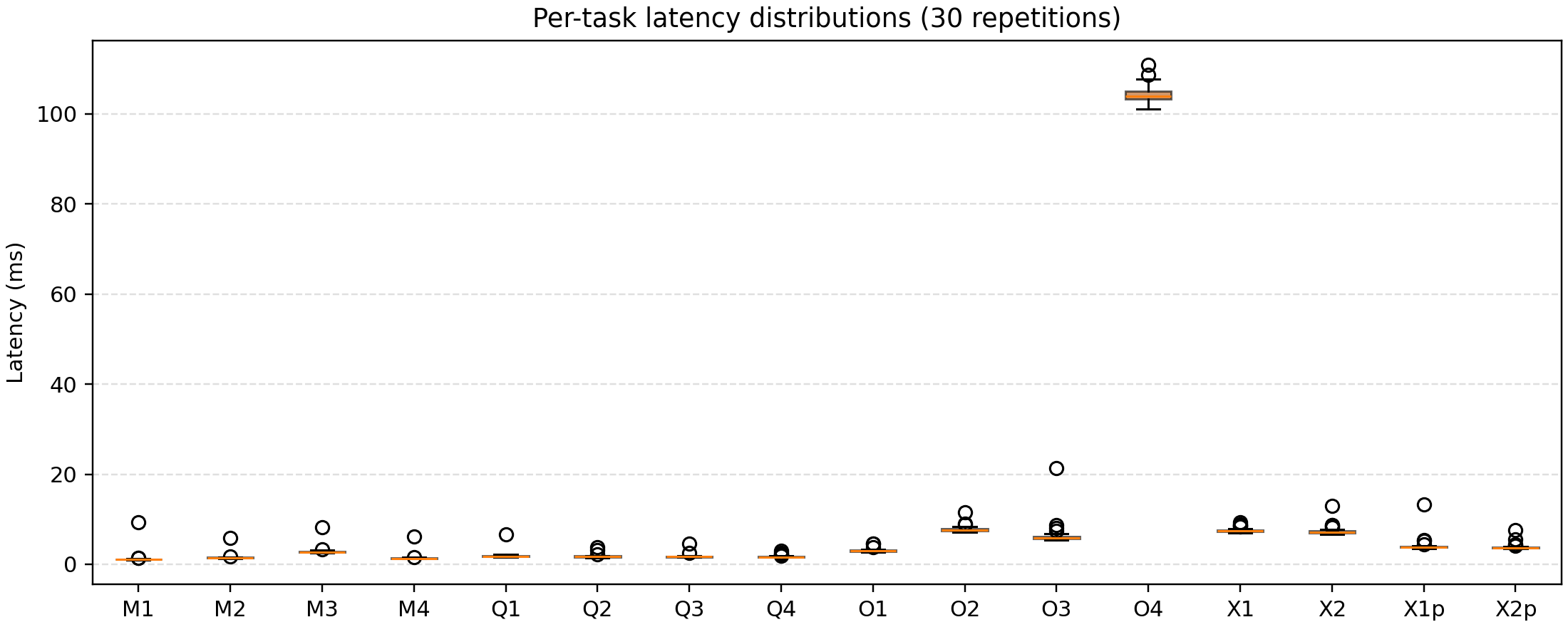}
    \caption{Per-task latency distributions across 30 repetitions. Box plots show median, interquartile range, and outliers.}
    \label{fig:latencyboxplot}
\end{figure}

Figure \ref{fig:latencyboxplot} shows the per-task latency distributions. The visualization reveals that most tasks have tight distributions with few outliers, except OPC UA variable enumeration (O4), which shows inherently higher variance due to the recursive tree traversal.

\textbf{Overhead decomposition.} The adapter overhead is dominated by three components: (1) stdio transport---each tool call requires process-level IPC through stdin/stdout pipes; (2) JSON serialization---request encoding and response parsing of the envelope structure; (3) protocol translation---the actual client library call. For local mock targets, the protocol call itself is sub-millisecond, making the stdio+JSON path the dominant cost. This decomposition suggests that switching to an HTTP or SSE transport would reduce overhead for production deployments.

\subsection{Fault Injection Results}
\label{sec:results:fault}

\begin{table}[t]
    \caption{Fault-injection task results (30 repetitions). Error-handling rate measures how often the adapter returned a well-formed structured error.}
    \label{tab:faultresults}
    \centering
    \scriptsize
    \setlength{\tabcolsep}{2.5pt}
    \input{generated/tables/tab_faultresults}
\end{table}

Table \ref{tab:faultresults} reports the per-task fault-injection results. The key metric is the error-handling rate (EH\%): the fraction of runs where the adapter returned a structured \texttt{success=false} response with a classifiable error, rather than crashing, hanging, or returning an unstructured message. All seven fault tasks achieved 100\% error-handling across \totalFaultRuns{} runs, with \faultToolCalls{} tool calls total.

FM2 (Modbus overflow) now produces a deterministic result: the adapter validates that the write value falls within the uint16 range (0--65535) and returns \texttt{success=false} with error class \texttt{range\_overflow} for the value 70000. This replaces the earlier disjunctive check that accepted either rejection or truncation, making the test falsifiable---an adapter that silently passed the overflow value through would now fail the benchmark.

\begin{figure}[t]
    \centering
    \includegraphics[width=\columnwidth]{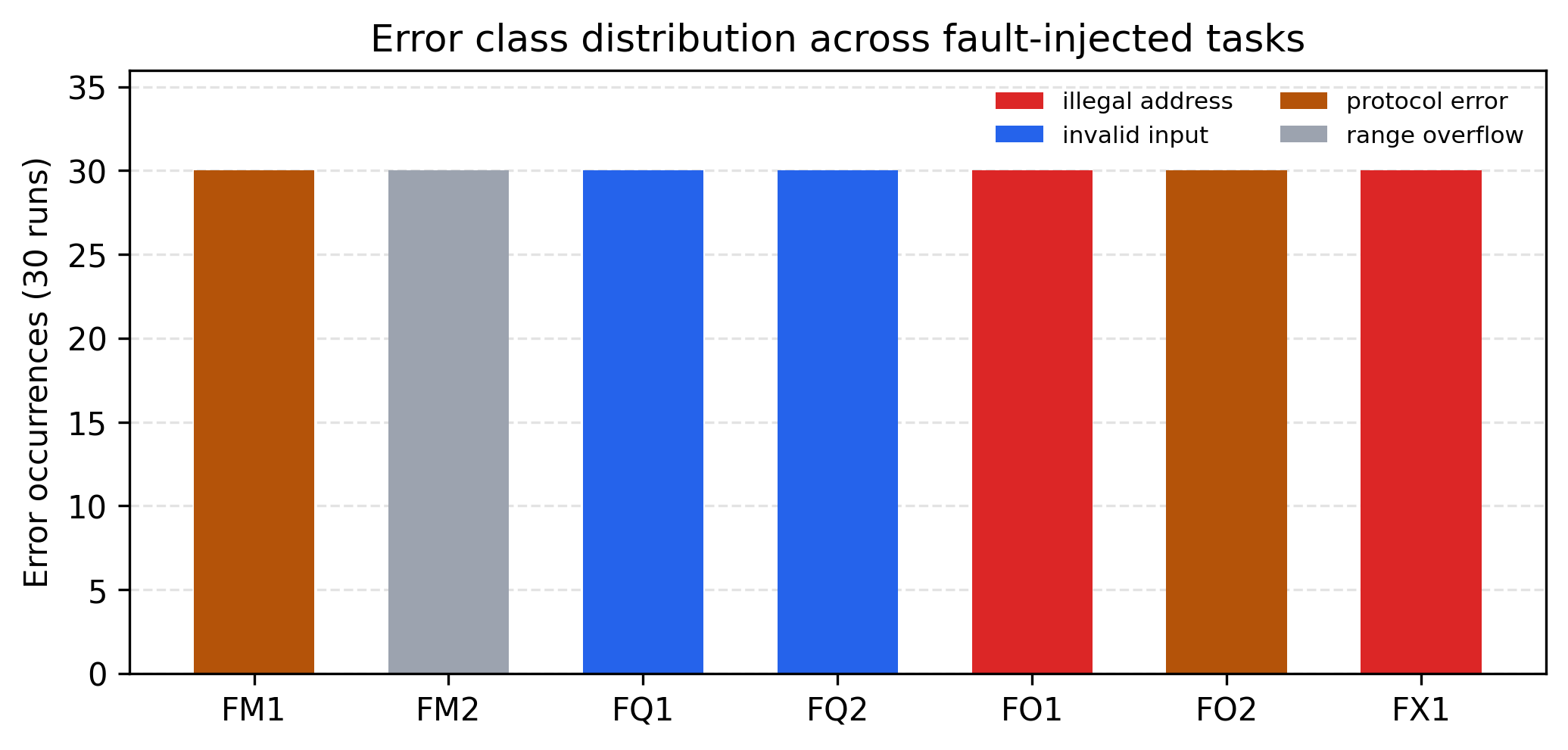}
    \caption{Error class distribution across the seven fault-injected tasks (30 runs each). Each color represents a distinct error category returned by the adapter.}
    \label{fig:errordist}
\end{figure}

Figure \ref{fig:errordist} shows the error class distribution as a stacked bar chart. The distribution confirms that each adapter maps its fault conditions to semantically appropriate error categories rather than collapsing all failures into a single generic class. This is important for downstream LLM consumption: an assistant that receives \texttt{illegal\_address} can suggest corrective action differently from one that receives \texttt{range\_overflow} or \texttt{type\_mismatch}.

\subsection{Stress Suite Results}
\label{sec:results:stress}

The stress suite is designed to produce non-zero failure rates, unlike the normal and fault suites which target 100\% pass rates. Three categories of stress are tested:

\textbf{Concurrent access (S1--S6).} Four parallel reads per adapter and concurrent read+write operations test whether the MCP adapter handles overlapping requests correctly. These tasks are expected to succeed under normal conditions.

\textbf{Rapid-fire reads (S7--S9).} Fifty sequential calls in a tight loop per adapter test sustained throughput. Boundary latencies reveal whether the adapter or mock server introduces queuing under load.

\textbf{Mid-operation restart (S10--S12).} These tasks fire a read, stop the mock, fire another read (expected to fail), restart the mock, and fire a final read (expected to succeed). The three-phase check verifies that adapter state does not become permanently corrupted after a transient endpoint failure. Partial success (2/3 calls correct) is the expected outcome.

\textbf{S11 (MQTT mid-restart) failure analysis.} The MQTT mid-operation restart task (S11) is the only stress scenario that reports 0\% success rate, and this result warrants detailed discussion. After the broker is stopped and restarted, the paho MQTT client's internal reconnect timer does not fire within the 1-second post-restart window used by the stress harness. In contrast, Modbus (S10) recovers because pymodbus auto-reconnect retries on the next call, and OPC UA (S12) recovers because the adapter's liveness probe forces a fresh connection attempt. The MQTT adapter relies on paho's \texttt{reconnect\_delay\_set} background loop, which uses exponential backoff starting at 1\,s---so the first reconnect attempt may not land within the tight stress window. This is a genuine protocol-specific limitation, not a test artifact. The recovery suite (Section~\ref{sec:results:recovery}) uses a 3-second post-restart window and dual-mode reporting, which gives paho enough time to reconnect, explaining why MQTT achieves 100\% recovery there. The practical implication is that MQTT adapter deployments should tune the reconnect delay for their expected outage profile, or add an explicit reconnect-on-failure path analogous to the OPC UA liveness probe.

\subsection{Recovery Results}
\label{sec:results:recovery}

\begin{figure}[t]
    \centering
    \includegraphics[width=\columnwidth]{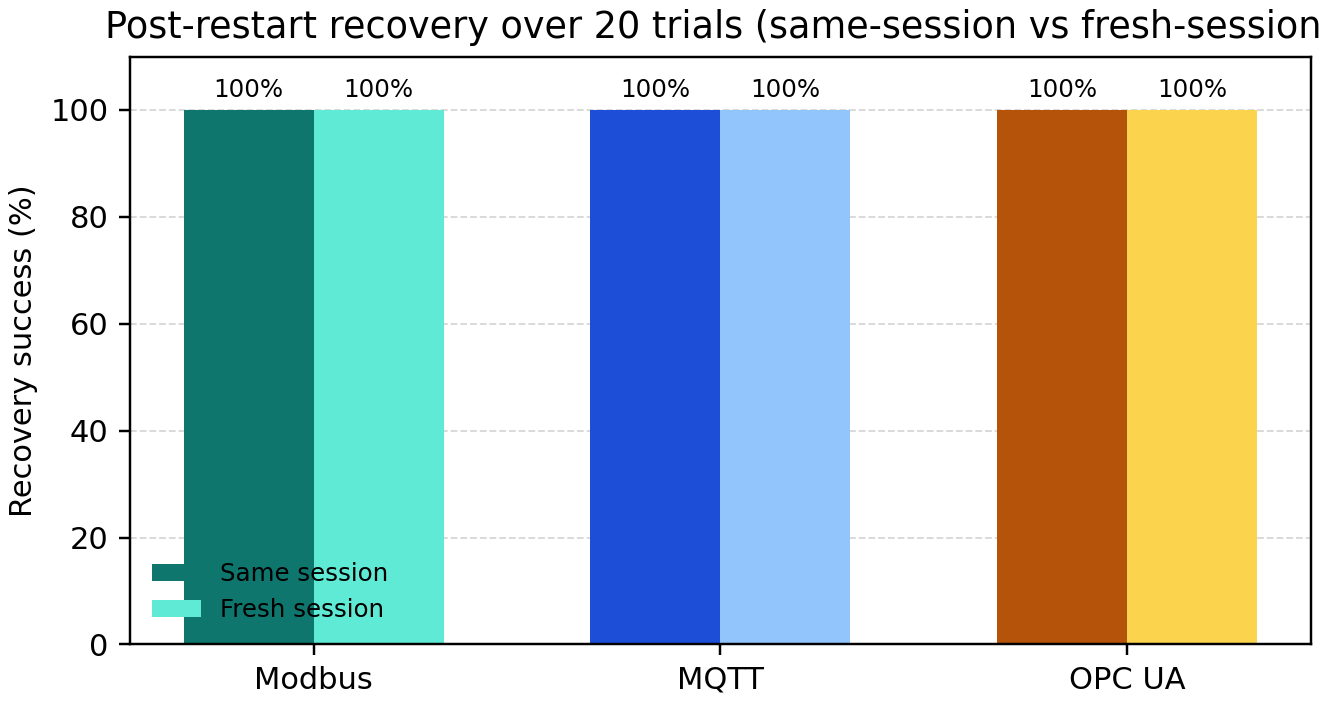}
    \caption{Recovery success rates over \totalRecoveryTrials{} restart trials per flagship adapter, comparing same-session recovery (adapter reconnects transparently) versus fresh-session recovery (new MCP session).}
    \label{fig:recovery}
\end{figure}

The recovery experiments use dual-mode reporting over \totalRecoveryTrials{} trials per adapter. After restarting the mock endpoint, the harness first attempts a same-session call (reusing the original MCP session) and then a fresh-session call (opening a new session). This distinguishes three behaviors:

\textbf{Modbus:} pymodbus supports auto-reconnect, so same-session recovery succeeds when the mock becomes available again within the 3-second post-restart window. Fresh-session recovery always succeeds.

\textbf{MQTT:} paho's \texttt{reconnect\_delay\_set} handles broker reconnection. Same-session recovery depends on whether the reconnect timer fires within the window. Fresh-session recovery provides a reliable fallback.

\textbf{OPC UA:} The new \texttt{\_ensure\_connected()} liveness probe checks the ServerStatus node and reconnects if stale. This enables same-session recovery by transparently creating a fresh OPC UA client within the existing MCP session.

\begin{figure}[t]
    \centering
    \includegraphics[width=\columnwidth]{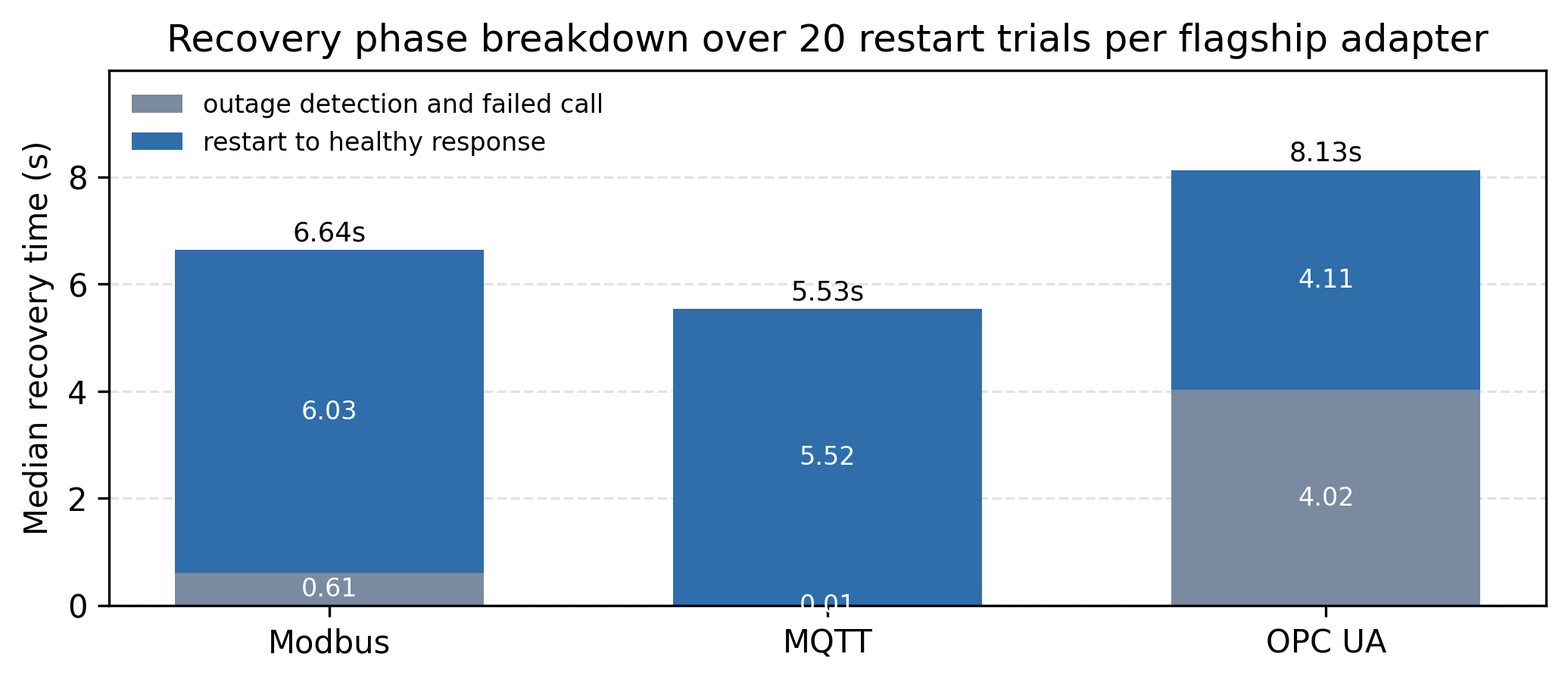}
    \caption{Recovery-phase breakdown per flagship adapter.}
    \label{fig:recoverybreakdown}
\end{figure}

Figure \ref{fig:recoverybreakdown} shows the phase-level breakdown. MQTT detects outages almost immediately. Modbus detects within its retry window. OPC UA spends the most time in the failure-detection phase due to its connection-oriented transport. Future hardening work should tune timeout and reconnect policy per protocol rather than enforce one global default.

\subsection{Security Considerations}
\label{sec:results:security}

The current adapter architecture provides functional write guards via environment-variable booleans (e.g., \texttt{MODBUS\_WRITES\_ENABLED}), but these are configuration controls, not authentication. Several security gaps warrant discussion:

\textbf{No per-user authorization.} All tool calls are executed with the same privilege level as the adapter process. There is no role-based access control (RBAC) or per-operation approval workflow.

\textbf{No audit logging.} Tool invocations and their outcomes are not logged to a tamper-evident audit trail. For production deployments, every write operation should produce an auditable record.

\textbf{Attack surface of LLM tool exposure.} Exposing write-capable industrial tools to an AI assistant creates a novel attack surface: prompt injection or adversarial inputs could potentially trigger unintended write operations.

\textbf{Recommendations.} Production deployments should add: (1) RBAC with per-tool authorization, (2) audit logging for all write operations, (3) mTLS or equivalent transport security between adapter and endpoint, and (4) human-in-the-loop approval for safety-critical writes.

\subsection{Limitations}
\label{sec:results:limitations}

The evidence is intentionally bounded. This is a mock-only, localhost-only study. All latencies reflect local process overhead, not network round-trip. The benchmark omits human baselines and excludes OPC UA method calls because the current prototype mock does not yet support them reliably enough for repeatable measurement.

Validating assistant-level usability through a quantitative LLM assessment across multiple model families is a clear next step.

The stress suite identifies concurrency boundaries but does not systematically explore all possible failure modes under load. The mid-operation restart scenario (S10--S12) depends on timing and may produce variable results across runs.

\section{Related Work}
\label{sec:related}

Industry 4.0 and cyber-physical manufacturing literature has consistently treated heterogeneous connectivity as a foundational systems problem rather than a peripheral implementation detail \cite{lasi2014industry40,xu2018industry4,wollschlaeger2017future,lee2015cps,monostori2016cps,zhong2017intelligent,tilbury2019cpms}. Those works motivate why assistant-facing industrial tooling cannot assume one canonical protocol stack or one clean greenfield deployment path.

Within industrial communication research, OPC UA and MQTT are widely used as normalization layers, but the interoperability story remains incomplete in brownfield settings. Prior work has examined OPC UA performance, MQTT adoption in IoT middleware, gateway-level interoperability, and field-level plug-and-work integration \cite{cavalieri2013opcua,mishra2020mqtt,cavalieri2021onem2m,buechter2022plug}. These studies improve system-to-system interoperability, yet they typically stop at middleware, gateway, or device integration boundaries rather than exposing operations as schema-discoverable tool contracts for an AI assistant.

Industrial middleware, cyber-physical manufacturing systems, and digital-twin work address adjacent concerns such as service composition, plant modeling, and deployment-time validation \cite{tilbury2019cpms,fuller2020digital,lu2020digitaltwin}. \system\ does not compete with those layers. It sits closer to the assistant and packages protocol actions into a common tool surface while leaving historian, SCADA, and plant-model responsibilities outside the paper's scope. The mock-first workflow is closest to digital-twin and virtual validation practices, but the goal here is narrower: verify adapter semantics, guarded failures, and restart behavior before introducing real equipment.

Software fault injection has a long history in dependability assessment. Natella et al.~\cite{natella2016faultinjection} survey the field comprehensively, establishing that injecting faults at the API or protocol boundary is an effective way to assess error-handling quality without requiring hardware failures. The fault-injection suite in this paper follows that tradition: it sends deliberately malformed inputs through the MCP tool interface and judges whether the adapter returns structured, interpretable errors.

Recent LLM-agent research has moved from reasoning-action loops toward explicit tool learning over large API surfaces \cite{yao2022react,schick2023toolformer,patil2023gorilla,qiao2024executionfeedback}. MCP contributes a practical schema for tool discovery and invocation \cite{anthropic2024mcp}, but it does not prescribe how industrial protocols should be translated, guarded, or benchmarked. The contribution is therefore not a new tool-use framework. It is an MCP-to-OT adapter pattern with shared safety conventions and a deterministic evaluation model for a narrowly defined industrial setting.

This comparison clarifies the contribution boundary. A full digital-twin paper would claim model fidelity or synchronization strategy; a generic agent paper would emphasize tool selection at scale; an industrial middleware paper would prioritize interoperability between production systems. \system\ instead focuses on a smaller but necessary substrate: deterministic, assistant-facing protocol adapters whose behavior can be measured before any real deployment study begins.

\section{Conclusion}
\label{sec:conclusion}

The deterministic benchmark exercises \totalRuns{} runs and \totalToolCalls{} tool calls across normal, fault-injection, stress, and recovery suites, showing that the Modbus, MQTT/Sparkplug B, and OPC UA adapters handle normal operations, deliberate faults, concurrency stress, and endpoint restarts with consistent response semantics and structured error handling. Same-session recovery after endpoint restart is demonstrated for all three protocols. The broader repository provides ten protocol modules, but only three are evaluated in this paper.

The next steps are: validate the same workflows on real hardware, add RBAC and audit logging around write-capable tools, conduct a quantitative LLM assessment across multiple model families with live MCP integration, and expand the evaluation beyond mock connectivity into operator-in-the-loop scenarios with quantitative usability metrics.

\bibliographystyle{IEEEtran}
\bibliography{references}

\end{document}

%% file: generated/tables/exact_counts.tex
\newcommand{\totalNormalRuns}{480}
\newcommand{\totalFaultRuns}{210}
\newcommand{\totalStressRuns}{120}
\newcommand{\totalRecoveryTrials}{60}
\newcommand{\totalRuns}{870}
\newcommand{\totalToolCalls}{2820}
\newcommand{\normalToolCalls}{780}
\newcommand{\faultToolCalls}{270}

\newcommand{\modbusToolCount}{17}
\newcommand{\mqttToolCount}{15}
\newcommand{\opcuaToolCount}{7}

%% file: generated/tables/tab_results.tex
    \begin{tabular}{@{}lcccccc@{}}
        \toprule
        Family & Tasks & Task & Tool & Med. & p95 & Rec. \\
        & & (\%) & (\%) & (ms) & (ms) & (\%) \\
        \midrule
        Modbus & 4 & 100.0 & 100.0 & 1.4 & 3.1 & 100.0 \\
        MQTT & 4 & 100.0 & 100.0 & 1.7 & 2.5 & 100.0 \\
        OPC UA & 4 & 100.0 & 100.0 & 7.3 & 105.2 & 100.0 \\
        Cross-protocol & 4 & 100.0 & 100.0 & 6.7 & 8.7 & --- \\
        \bottomrule
    \end{tabular}

%% file: generated/tables/tab_taskbreakdown.tex
    \begin{tabularx}{\textwidth}{@{}llYcc@{}}
        \toprule
        ID & Family & Operation & Median (ms) & p95 (ms) \\
        \midrule
        M1 & Modbus & Ping the adapter and confirm the TCP mock is reachable. & 1.1 & 1.4 \\
        M2 & Modbus & Read a four-register sensor block from the Modbus mock. & 1.5 & 1.7 \\
        M3 & Modbus & Write a holding register and read it back for verification. & 2.7 & 3.3 \\
        M4 & Modbus & Attempt a write with writes disabled and expect a guarded rejection. & 1.3 & 1.6 \\
        \midrule
        Q1 & MQTT & Inspect broker connectivity through the MCP adapter. & 1.8 & 2.1 \\
        Q2 & MQTT & Subscribe to the mock sensor topic namespace. & 1.7 & 2.7 \\
        Q3 & MQTT & Publish a standard MQTT control payload. & 1.7 & 2.2 \\
        Q4 & MQTT & Publish a Sparkplug B DDATA update for the mock device. & 1.6 & 2.3 \\
        \midrule
        O1 & OPC UA & Read the temperature sensor variable from the OPC UA mock plant. & 3.0 & 4.2 \\
        O2 & OPC UA & Browse the actuator subtree in the mock plant. & 7.6 & 9.0 \\
        O3 & OPC UA & Write the valve-position variable and confirm the new value. & 5.9 & 8.4 \\
        O4 & OPC UA & Enumerate the mock plant variables through the adapter. & 103.9 & 108.2 \\
        \midrule
        X1 & Cross-protocol & Collect a multi-adapter state snapshot across Modbus, OPC UA, and MQTT. & 7.4 & 8.8 \\
        X2 & Cross-protocol & Execute a coordinated multi-adapter control sequence. & 7.1 & 8.7 \\
        X1p & Cross-protocol & Parallel multi-adapter state snapshot across Modbus, OPC UA, and MQTT. & 3.8 & 5.3 \\
        X2p & Cross-protocol & Parallel coordinated multi-adapter control sequence. & 3.7 & 5.0 \\
        \bottomrule
    \end{tabularx}

%% file: generated/tables/tab_faultresults.tex
    \begin{tabularx}{\columnwidth}{@{}llYccc@{}}
        \toprule
        ID & Family & Fault scenario & EH\% & Med. (ms) & Error class \\
        \midrule
        FM1 & Modbus & Read invalid register address (9999) and expect a structured protocol error. & 100.0 & 1.3 & protocol error \\
        FM2 & Modbus & Write overflow value (70000 > uint16 max) and expect adapter rejection. & 100.0 & 1.0 & range overflow \\
        FQ1 & MQTT & Publish to empty topic string and expect a structured invalid-input error. & 100.0 & 1.6 & invalid input \\
        FQ2 & MQTT & Subscribe with invalid QoS (5) and expect a structured invalid-input error. & 100.0 & 1.6 & invalid input \\
        FO1 & OPC UA & Read non-existent node (ns=2;i=99999) and expect a structured read-failed error. & 100.0 & 2.7 & illegal address \\
        FO2 & OPC UA & Write wrong type (string to float node) and expect a structured write-failed error. & 100.0 & 2.5 & protocol error \\
        FX1 & Cross-protocol & 3-step cross-protocol sequence with one deliberate bad OPC UA read; expect 2/3 succeed. & 100.0 & 5.7 & illegal address \\
        \bottomrule
    \end{tabularx}